\documentstyle[12pt,rotating]{article}
\newcounter{popnr}
\addtocounter{popnr}{1}

\newcommand{\beq}{\begin{eqnarray}}
\newcommand{\eeq}{\end{eqnarray}}
\newcommand{\beqq}{\begin{eqnarray*}}
\newcommand{\eeqq}{\end{eqnarray*}}
\begin{document}

\setcounter{page}{0}
\thispagestyle{empty}
\begin{titlepage}

\title{Second Loop Corrections from Superheavy Gauge Sector to Gauge
Coupling Unification}
\author{{\em Dariusz Grech}\\
\\
Institute of Theoretical Physics\\
University of Wroc{\l}aw, Maks Born Sq.9,\\
 PL-50-204 Wroc{\l}aw, POLAND}
\date{}
\maketitle

\begin{abstract}
We deal with extensions of the Standard Model (SM) adding horizontal 
interactions
between particle generations. We calculate two loop corrections 
caused 
by the presence 
of coupling between hypothetical horizontal gauge bosons and matter
field at high energy.
It is shown that coupling of such extra bosons does not affect up to
two loop level the
positive features of unified and extended SM with horizontal 
symmetry discussed in former publications. Corrections from bosonic 
horizontal sector make about tenth part of those caused by fermionic
sector.
Although small they are however larger than accuracy of some 
electroweak 
measurements and therefore they might be important for future
verification of various proposed horizontal models.
\end{abstract}

\end{titlepage}
\section{Introduction}
Standard model of strong and electroweak interactions (SM) based on
$SU(3)_C\\
\times SU(2)_L \times U(1)_Y$ gauge group has got many 
unquestionable achievments. Ne\-ver\-the\-less it does not seem to be 
complete. 
Everyone agrees it contains too
many arbitrary parameters. Their origin can in general be explained
only in larger schemes which incorporate additional global or local
symmetries. Such models beyond the SM can be divided into SUSY models
- with Minimal Supersymmetric Standard Model (MSSM) as the best
candidate,
or  schemes with extended gauge sector. In the latter type of 
theories 
the $SU(3)_C \times SU(2)_L \times
U(1)_Y$ gauge group above the electroweak mass scale $M_Z$ is extended
to $SU(3)_C \times SU(2)_L \times
U(1)_Y \times G$ group product where $G$ is new exact gauge symmetry
broken at some
mass scale $M_G>M_Z$. Most popular left-right symmetric models and
horizontal symmetry models belong to such class of theories.
One should note that junctions between these models and SUSY can also 
be considered. How far will
fermions and bosons associated with $G$ change or explain parameters 
of
SM?

We will deal here with models in which $G$ is taken as horizontal 
gauge
group $G_H$. Analysis of such models was first
performed in ref. \cite{1}. Two loop approach shows that
electroweak parameters are significantly changed in such models in
"proper" direction from the experimental point of view [1]. 
Even for the large class of non-supersymmetric models the electroweak
mixing angle $\theta_W$ can be increased to $\sin^2\theta_W(M_Z ) \sim
0.23$. Simultaneously the origin of various fermionic generations can
be explained, and moreover,
if one unifies a model into a single simple gauge group according to
GUT scheme, the reasonable proton lifetime $\tau_p >\sim 10^{33}$ yrs
is to be obtained.

To solve the problem completely we need to take into account on two
loop level not only fermions associated with $G_H$ but also
intermediate gauge bosons generated by this group. An influence of
these bosons was not discussed in details in former publications,
where only preliminary results about horizontal gauge boson
sector were given. The complete analysis of this problem is a task 
we undertake now.

In order to focus investigation on some class of models ($G_H$ might 
in
general be arbitrary) we will force $G_H$ to satisfy few
conditions. The detailed list of all requirements is given in 
\cite{1}.
We will quote the most important ones: $SU(3)_C \times SU(2)_L \times
U(1)_Y \times G_H$ should be possible to be unified into simple gauge
group $G_{\small \rm GUT}$ with complex representation $\phi_{\small
\rm
GUT}$; $G_H$ is also assumed to be simple group and unification of
flavors within single generation should be permitted at some mass 
scale
$M_X$ before grand unification of flavors and generations occurs at
scale $\mu_X >M_X$.

Corrections caused by loops involving horizontal bosons will be
calculated from two loop solution of renormalization group equation 
for
effective coupling constants $\alpha_{\mu} \ (\mu=1,2,3,4)$ of 
$U(1)_Y,
\ SU(2)_L,\ SU(3)_C$ and $G_H$ respectively:
\beq
\alpha^{-1}_{\mu} (q^2) - \alpha^{-1}_{\mu} (Q^2)=(4\pi
)^{-1}\beta_{\mu}\ln \frac{q^2}{Q^2}  + (4\pi)^{-1}\sum_{\nu =1}^4
\frac{\beta_{\mu\nu}}{\beta_{\nu}} \ln
\frac{\alpha^{-1}_{\nu}(q^2)}{\alpha^{-1}_{\nu}(Q^2)}
\eeq
where $q^2, Q^2$ are momenta and $\beta_{\mu}, \beta_{\mu\nu}$- first
and second order $\beta$-functions. It is of particular interest to
estimate mixed terms with $\beta_{i4}\equiv \beta^H_i (i=1,2,3)$
responsible for two loop corrections involving horizontal bosons as 
well as
terms with $\beta_4 \equiv \beta^H$. We will perform calculations for
the class of
theories satisfying  the conditions quoted above. Such models were
found in \cite{1} and are listed once more in first columns of Table
1 with
the following data: representation structure $\phi_{\small\rm GUT}$ in
Young tableau language, the maximal number of fermionic generations
$n_G$,
and the complex part of fermionic representation after the breaking
$$
G_{\small \rm GUT}\stackrel{\mu_X}{\longrightarrow} G_f (\equiv
SU(5))\times G_H
$$
($G_f$ assumed here to be  $SU(5)$ is the unifying group of a single
generation).

Using general formulas given in ref.\cite{2} for the first and second
order
$\beta$-functions we arrive at
\beq
\beta^H=\frac{1}{3}\left\{\frac{11}{2}l(V^H)-\sum_k n_k l
(\psi^H_k)\right\}
\eeq
where $V^H$ is the adjoint representation of $G_H$, 
$\psi^H\equiv\sum_k
n_k \psi^H_k$ is decomposition of the complex part of $\phi_{GUT}$ 
into
irreducible representations $\psi^H_k$ of $G_H$, and $l(\phi)$ is
Dynkin index \cite{3} for representation $\phi$.

The second order $\beta_{\mu\nu}$ functions for $\mu\ne\nu$ read
\beq
\beta_{\mu\nu}=-l(\psi^{\mu})C_2(\psi^{\nu})
\eeq
where $C_2 (\psi^{\nu})$ is the Casimir eigenvalue for representation
$\psi^{\nu}$ of the group $G^{\nu}$. It can be expressed by the number
of free parameters of $G^{\nu}$ ($dim \ \ G^{\nu}$) and dimension 
of $\phi^{\nu}$
($dim \ \phi^{\nu}$):
\beq
C_2 (\phi^{\nu})=\frac{1}{2}l(\phi^{\nu})\frac{dim\ G^{\nu}}{dim\
\phi^{\nu}}
\eeq
By $\psi^{\mu (\nu )}$ in (3) we will understand nontrivial complex
part of $\phi_{GUT}\longrightarrow(\psi^{\mu}, \psi^{\nu})$ after
decomposition
$G_{GUT}\longrightarrow G^{\mu} \times G^{\nu} \
(\mu, \nu=1,2,3,4)$.

Substitution of (4) to (3) gives
\beq
\beta_{\mu\nu}=-\frac{1}{2}\sum_{\psi\in\phi_{GUT}}l(\psi^{\mu})l
(\psi^{\nu})\frac{dim\
G^{\nu}}{dim\ \psi^{\nu}}
\eeq
The general form of the complex part of $\phi_{GUT}$ after
decomposition
$G_{GUT}\stackrel{\mu_X}{\longrightarrow}G_f \times G_H$ will be
\beq
\phi_{GUT}\stackrel{\mu_X}{\longrightarrow}\sum_k n_k ({\bf s}_k,{\bf
h}_k)
\eeq
where ${\bf s}_k$ are fermionic representations of $G_f$, ($G_f
\stackrel{M_X}{\longrightarrow} SU(3)_C \times SU(2)_L \times U(1)_Y$)
and ${\bf h}_k$ are fermionic representations of $G_H$.
Therefore we may write
\beq
\beta^H_i =-\frac{1}{2} dim\ G_H \sum_k \frac{n_k l ({\bf s}_k)l({\bf
h}_k )}{dim\ {\bf h}_k}
\eeq
Let us note that the above formula does not depend on $i=1,2,3$. It
reflects the fact that unification within single generation 
(unification of flavours)
occurs
independently on the unification between distinct generations. Thus
\beq
\beta^H_3 =\beta^H_2 =\beta^H_1
\eeq
and therefore two loop contributions from horizontal bosons are the
same for all interactions.
The values of $\beta^H$ and $\beta^H_i$ are also shown in Table 1.

Now we may proceed to evaluate corrections according
to equation (1). It will be useful to keep in mind the schematic plot
of effective couplings versus $t=\ln q^2$ as shown in Fig.1.

\section{Corrections to the unifying mass $M_X$}
Comparing $\mu=3$ and $\mu=2$ terms in (1) one finds that if 
horizontal
boson coupling is switched off, i.e. $\beta^H_i \equiv 0$ then
\beq
\begin{array}{ll}
\alpha^{-1}_3 (M_H)-\alpha^{-1}_2(M_H)=(4\pi )^{-1} (\tilde\beta_3
-\tilde\beta_2 )\ln \frac{M_H^2}{M_X^2}+&\null\\
+(4\pi )^{-1}\sum^3_{j=1}\frac{\tilde\beta_{3j} -
\tilde\beta_{2j}}{\tilde\beta_j} \ln \frac{\alpha^{-1}_j (M_H
)}{\alpha^{-1}_j (M_X )}
\end{array}
\eeq
where $\tilde\beta$ represents $\beta$-functions for $q^2>M_H^2$ and 
$\alpha_3 (M_X) =\alpha_2 (M_X) \equiv \alpha_G(M_X) $ is assumed for
$\alpha_G$ being the unified coupling constant of $G_f$

Similarly if one switches horizontal boson coupling on i.e. $\beta^H_i
\neq 0$ :
\beq
\begin{array}{ll}
\alpha^{-1}_3 (M_H )-\alpha^{-1}_2 (M_H ) =&(4\pi )^{-1}(\tilde\beta_3
- \tilde\beta_2 ) \ln \frac{M^2_H}{\tilde M^2_X} +\null\\ \\
\null &\null+(4\pi )^{-1}
\sum^4_{\nu =1}\frac{\tilde\beta_{3\nu}
-\tilde\beta_{2\nu}}{\tilde\beta_{\nu}} \ln \frac{\alpha^{-
1}_{\nu}(M_H
)}{\alpha^{-1}_{\nu}(\tilde M_X )}
\end{array}
\eeq
where $\tilde M_X$ is the unifying mass corrected by the presence of
such coupling.

Subtracting (9) from (10) we arrive with
\beq
\ln \frac{\tilde M_X}{M_X} =\frac{1}{2\tilde\beta^H}
\frac{\tilde\beta^H_{3} - \tilde\beta^H_{2}}{\tilde\beta_3
-\tilde\beta_2} \ln \frac{\alpha_H (M_X )}{\alpha_H (M_H )}
\eeq
for $\alpha_H = \alpha_4$. 
Obviously $\tilde\beta^H = \beta^H$, $\tilde\beta^H_{i} = 
\beta^H_{i}$ 
and hence, because of (8), $\tilde M_X =M_X$.
This result was obvious anyway; the corrections from $\beta^H_{i}$
shift,
according to (8), all plots $\alpha_i (q^2 )$ in Fig.1
by the same amount leaving $M_X$ at which intersection occurs -
unchanged.
We see that corrections from gauge boson sector of
$G_H$ do not change $M_X$ up to two loop level.
This means also that the  expected proton
lifetime $\tau_p \sim M^4_X$ should not be changed.

\section{Corrections to the horizontal mass scale {\bf $M_H$}}
$G_H$ is broken at $M_H$ and this mass scale changes the
behaviour of effective coupling constants with energy  (see Fig.1). 
This fact is
important in estimation of $\sin ^2 \theta_W (M_Z )$ expressed by
$\alpha_2 (M_Z )$ and $\alpha_1 (M_Z )$ according to the formula
\beq
\sin^2 \theta_{W} (M_Z ) = \left ( 1+\frac{5}{3} \frac{\alpha_2
(M_Z )}{\alpha_1 (M_Z )} \right ) ^{-1}
\eeq
Uncorrected $M_H$ values calculated previously for various admissible
models are quoted in the sixth column of Table 1. (for details of such
calculation see [1]). Now we want to estimate corrections due to the
presence of extra bosonic sector of $G_H$. Let us remark we have two 
regions of
energy: one with $q^2 <M_H^2$ and the second with $q^2 >M_H^2$. All
$G_H$ 
bosons as well as
those fermions of $\phi_{GUT}$ which transform with respect to real
representations of $G_f$ (or $SU(3)_C \times SU(2)_L \times U(1)_Y $)
are assumed to decouple below $M_H$. Respective $\beta$-functions in 
these two
regions will be denoted by $\beta$ ($q^2 < M^2_H $) and $\tilde\beta$
($q^2 > M^2_H$ ).

Subsequently we get from (1) for $\beta^H_i \equiv 0$:
\beq
\begin{array}{ll}
4\pi \Delta \alpha^{-1}_3 &\stackrel{\rm def}{=} 4\pi (\alpha^{-1}_3
(M_Z)-\alpha^{-1}_3 (M_X ))=\\
\null&=(\tilde \beta_3 -\beta_3 )\ln M^2_H +\beta_3 \ln M^2_Z
-\tilde\beta_3 \ln M^2_X + \sum^3_{j=1}
\frac{\beta_{3j}}{\beta_j}\ln\frac{\alpha^{-1}_j (M_Z )}{\alpha^{-1}_j
(M_H )}\\
\null&\null
-\sum^3_{j=1}\frac{\tilde\beta_{3j}}{\tilde\beta_j}\ln
\frac{\alpha^{-1}_j(M_X )}{\alpha^{-1}_j (M_H )}
\end{array}
\eeq
Both terms on the left-hand side of (13) are fixed; $\alpha_3 (M_Z )$
is
the experimental input while $\alpha_3 (M_X ) \equiv \alpha_G (M_X )$
is
to be calculated to ensure perturbativity of a theory near
$M_X$ ([1]). Results for $\alpha_G (M_X )$ obtained numerically 
are also quoted for
reference in Table 1.

Hence if horizontal boson sector is switched on one should get
\beq
\begin{array}{ll}
4\pi\Delta\alpha^{-1}_3 &=4\pi (\alpha^{-1}_3 (M_Z ) -\alpha^{-1}_3
(\tilde M_X ))=\\
\null&=(\tilde\beta_3 -\beta_3 )\ln \tilde M^2_H + \beta_3 \ln M^2_Z -
\tilde\beta_3 \ln \tilde M^2_X + \sum^3_{j=1} 
\frac{\beta_{3j}}{\beta_j}
\ln\frac{\alpha^{-1}_j (M_Z )}{\alpha^{-1}_j (M_H )} \\
\null&\null - \sum^3_{j=1}\frac{\tilde\beta_{3j}}{\tilde\beta_j}
\ln\frac{\alpha^{-1}_j (\tilde M_X )}{\alpha^{-1}_j (\tilde M_H
)}-\frac{\tilde\beta^H_3}{\tilde\beta^H} \ln\frac{\alpha^{-1}_H 
(\tilde
M_X )}{\alpha^{-1}_H (\tilde M_H )}
\end{array}
\eeq
where $\tilde M_H$ is the corrected value of $M_H$.\\
Comparing (13) with (14) and keeping $ M_X \sim \tilde M_X, 
\alpha_j(M_X) = \alpha_j(\tilde M_X)
= \alpha_G(M_X)$ we obtain:
\beq
\frac{\tilde M_H}{M_H} = \left (\frac{\alpha_H (M_H )}{\alpha_H (M_X
)}\right )^{\frac{\beta^H_3}{2\beta^H (\tilde\beta_3 -\beta_3 )}}
\eeq
To use the above formula we need to know the ratio $\frac{\alpha_H 
(M_H
)}{\alpha_H (M_X )}$. This ratio will also be crucial in next section 
to determine
the corrections to $\sin^2 \theta_W$.\\
We receive on one loop level
\beq
\frac{\alpha_H (M_X )}{\alpha_H (M_H )} =1+(4\pi )^{-1}\beta^H\alpha_H
(M_X )\ln \left ( \frac{M_H}{M_X}\right ) ^2
\eeq
where $\alpha_H (M_X )$ should be estimated. Evolution of $\alpha_H$
and
$\alpha_G$ for $M^2_X <q^2 < \mu^2_X$ indicates (see Fig. 1) that
\beq
\alpha^{-1}_H (M_X ) - \alpha^{-1}_{GUT} (\mu_X ) = (4\pi )^{-
1}\beta^H
\ln\frac{M^2_X}{\mu^2_X}
\eeq
\beq
\alpha^{-1}_G (M_X ) -\alpha^{-1}_{GUT} (\mu_X ) = (4\pi )^{-1} 
\beta_5
\ln\frac{M_X^2}{\mu^2_X}
\eeq
where $\alpha_{GUT} (\mu_X )$ is the coupling constant of $G_{GUT}$ at
$\mu_X$ and $\beta_5$ is the one loop $\beta$-function for 
$G_f=SU(5)$ 
gauge theory.

Subtraction (17) from (18) yields to
\beq
\alpha_H (M_X ) = [\alpha^{-1}_G (M_X ) + (2\pi )^{-1}
(\beta_5 -\beta^H)\ln \frac{\mu_X}{M_X} ]^{-1}
\eeq
All terms above are known except $\mu_X$. To stay in perturbative
region we need to satisfy $\alpha^{-1}_{GUT} (\mu_X ) >1$. Therefore
\beq
\frac{\mu_X}{M_X}\leq\kappa
\eeq
where $\kappa$ is estimated with the help of (18):
\beq
\kappa = \exp\{\frac{2\pi}{|\beta_5|}(\alpha^{-1}_G -1)\}
\eeq
(Note that $\beta_5 < 0$ ).\\
Values of $\kappa$ estimated this way are also collected in Table 1.
Let us remark that in all cases $\mu_X$ does not exceed the Planck
mass $M_P \sim 10^{19} GeV$! Formulas (19)-(21) after substitution
to (15)-(16) allow to calculate the relative correction
$\tilde M_H /M_H$. We notice from (15), (16) and (19) it is maximal
if unification of flavors within generation
occurs at the similar scale as the unification of generations 
i.e. $\mu_X = M_X$.
In that case corrections $\delta M_H/M_H$, where 
$\delta M_H\stackrel{\rm def}{=}\tilde M_H - M_H$ are
tabularised (see Table 1). Their exact values depend 
strongly on the chosen model
and vary from $2\%$ to $12\%$ of its initial value. Thus the presence
of bosonic horizontal sector slightly increases the value of $M_H$.

\section{Corrections to $\sin^2 \theta_W$.}
Finally we approach the most intriguing question which concerns
corrections to
$\sin^2 \theta_W (M_Z )$. Formula (12) indicates that all changes in
$s\equiv \sin^2 \theta_W (M_Z )$  are due to changes in $\alpha_1 (M_Z
)$ and $\alpha_2 (M_Z )$. These couplings can be calculated from the
value $\alpha_G (M_X )$ using formula (1) and assuming the
step-function behaviour of $\beta$-function near the $M_H$ threshold.
One easy finds this way:
\beq
\begin{array}{ll}
\delta\alpha^{-1}_1(M_Z) \stackrel{\rm def}{=}&\alpha^{-1}_1 
(\beta^H_1 \neq 0)
-\alpha^{-1}_1 (\beta^H_1 =0) = (4\pi )^{-1}\frac{\beta^H_1}{\beta^H}
\ln \frac{\alpha^{-1}_H (M_H )}{\alpha^{-1}_H (M_X )}
\end{array}
\eeq
Due to (8) we have:
\beq
\delta\alpha^{-1}_2 (M_Z)\stackrel{\rm def}{=}\alpha^{-1}_2 
(\beta^H_2 \neq 0 ) -
\alpha^{-1}_2 (\beta^H_2 =0) = \delta\alpha^{-1}_1
\eeq
Numerical estimation of $\delta\alpha^{-1} \equiv \delta\alpha^{-1}_1
=\delta\alpha^{-1}_2$ is straightforward if one uses expressions (15)
and (16) as the input.

Now we proceed to calculate $\delta s$ - the correction to 
$\sin^2 \theta_W$ -
in terms of $\delta\alpha^{-1}$.
Differentiating (12) with respect to $\alpha^{-1}_1 = \frac{3}{5}
(1-s)\alpha^{-1}_0$ and $\alpha^{-1}_2 =s\alpha^{-1}_0$, where
$\alpha_0 =1/128$ is the electromagnetic coupling constant at $M_Z$
we get
\beq
\delta s = \alpha_0 \delta\alpha^{-1} \left(1 - \frac{8}{3} s\right)
\eeq
Maximal corrections correspond to $\mu_X = M_X$ and they are 
collected 
for all models at the end
of Table 1. They are typically of order $\sim 10^{-3}$ and still
increase $sin^2 \theta_W$ up to tenth part of second loop corrections
to
$sin^2 \theta_W$ caused by  horizontal fermionic sector itself. Indeed
as it was shown in ref.[1] fermionic sector in horizontal models is
able
on its own to increase $sin^2 \theta_W$ from $0.216$ to $0.233$. These
values are quoted for reference in Table 1. Let us notice that second
loop 
horizontal bosons corrections, although small, can already be larger
than
experimental accuracy in measurements of electroweak angle. Actually
all LEP
results and specific SLAC collider data give at present the world
average
value $sin^2 \theta_W = 0.23156 \pm 0.00020$ [4]. Therefore these
corrections
may play soon a crucial role for acceptance or rejection various 
proposed models.
It is already seen that $SU(2)_H$ is too small as horizontal 
gauge group candidate
predicting $sin^2 \theta_W$ slightly below the quoted value.\\
Let us finally comment on the possibility of massive neutrinos in 
presented models.
All of them contain in its $\phi_{GUT}$ singlet part of $G_f$ 
which makes space
for righthanded neutrinos $\nu_R$. Therefore mass terms for 
Dirac type $\nu$'s
can be obtained. Table 1 shows possible representation space of 
$\phi_{GUT}$
in which $\nu_R$'s can be embedded.\\

Summarizing obtained results one concludes that extra gauge bosons 
related to additional gauge group $G_H$ in
theories beyond the SM do not remarkably change the evolution of gauge
couplings of $SU(3)_C \times SU(2)_L \times U(1)_Y$
but positive features of these models 
will be even magnify if extra gauge bosons coupling is turned on.\\

\vspace{1.0 cm}
Author wishes to thank ICTP in Trieste
for his stay there, during which the part of this paper was done.

%

\unitlength 0.3mm
\begin{picture}(450,400)
\put(0,20){\vector(1,0){450}}
\put(0,20){\vector(0,1){260}}
\put(5,270){$\alpha (t)$}
\put(430,5){$t$}
\put(27,5){$M_Z$}
\put(227,5){$M_H$}
\put(317,5){$M_X$}
\put(377,5){$\mu_X$}
\put(30,20){\circle*{3}}
\put(230,20){\circle*{3}}
\put(320,20){\circle*{3}}
\put(380,20){\circle*{3}}
\put(30,60){$\alpha_1$}
\put(30,180){$\alpha_2$}
\put(30,255){$\alpha_3$}
      \put(300,165){$\alpha_H$}
\put(320,225){$\alpha_G$}
\put(410,225){$\alpha_{GUT}$}
\put(30,52){\line(4,1){200}}
\put(30,170){\line(4,-1){200}}
\put(30,250){\line(2,-1){200}}
\put(230,102){\line(5,6){90}}
\put(230,120){\line(1,1){90}}
\put(230,150){\line(3,2){90}}
\put(380,240){\line(-4,-5){140}} 
\put(320,210){\line(2,1){60}}
\put(380,240){\line(1,-1){70}}
\end{picture}

\noindent 
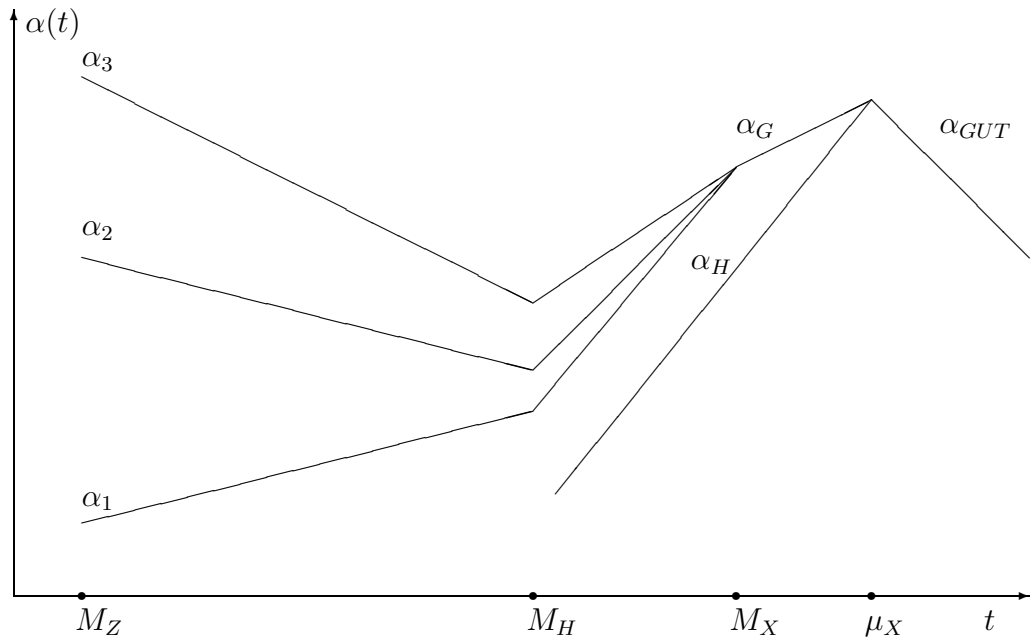
{Figure 1: Schematic plot of running coupling constants versus energy
$t=\ln q^2$ of strong, electroweak and horizontal interactions unified
together at the mass scale $\mu_X$ in the framework of $GUT$.}

\begin{figure}
\vspace*{-4cm}
\rotatebox{90}
{
{\scriptsize
$
\begin{array}{|c|c|c|c|l|l|l|l|l|l|l|l|l|l|l|}
\hline
G_{GUT}&G_H&\phi_{GUT}&n_G&G_f\times G_H \hbox{\rm complex part}&M_H
(GeV)&M_X (GeV)&\sin^2
\theta_W&\alpha_G&\beta^H&\beta^H_i&\kappa=\frac{\mu_X}{M_X}
&{\delta 
M_H
\over M_H} (\% )&G_f\times G_H \ \nu_R
{\rm space}&\delta\sin^2 \theta_W \\
\hline
SU(7)&SU(2)&6\{\bar{1}\}+2\{1^2\}+6\{1^3\}&4&2(5^*,2)+2(10,2)+\null&
4.4 \times 
10^9&1.21\times 10^{15}&0.228&0.3&-16&-15&133&7&2(1,1)+6(1,2)
&5\times 10^{-4}\\
&&&&&&&
& &&&&
&&
\\
&&&&\null+4(10,2)+8(10^*,1)&&&&&&&&&&\\
SU(8)&SU(3)&3\{\bar{1}\}+3\{\bar{1^2}\}+3\{1^3\}
&3
&(5^*,3^*)+(10,3)
&1.5\times
10^9
&1.03\times 10^{15}&0.230
&0.3
&-9
&-20
&55
&12
&3(1,1)
&1\times 10^{-3}\\
&&&&
+2(5^*,3^*)+2(10,3)
&&&&&&&&
&&\\
&&&&
+6(10^*,1)+3(5^*,1)
&&&&&&&&&&
\\
&&&&
+3(5,3^*)
&&&&&&&&
&&\\
SU(9)&SU(4)&19\{\bar{1}\}+3\{1^2\}+\{1^3\}+\{\bar{1^4}\}&4&(5^*
,4)+(10,4)+\null&1.3\times 10^{11}&1.32\times
10^{15}&0.233&0.3&-22&-115/4&3.8&7&(1,1)+3(1,6)&7\times 10^{-4}\\
&&&&\null+18(5^* ,1)+(5,6)+\null&&&&&&&&&&\\
&&&&\null+3(5,4)+20(1,4^* )+\null&&&&&&&&&&\\
&&&&\null+2(10,1)+(10,4^* )+\null&&&&&&&&&&\\
&&&&\null+(10^*,6)&&&&&&&&&&\\
SU(9)&SU(4)&13\{\bar{1}\}+2\{1^3\}+\{\bar{1^4}\}
&4
&(5^*,4)+(10,4)+\null
&1.3\times 10^{11}&1.25\times
10^{15}&0.234&0.3&-46/3&-125/4&3.8&9&(1,1)&1\times 10^{-3}\\
&&&&\null+2(5,6)+12(5^*,1)+\null&&&&&&&&&&\\
&&&&\null+15(1,4^*)+(10,4)+\null&&&&&&&&&&\\
&&&&\null+(10,4^*)+(10^*,6)+\null&&&&&&&&&&\\
&&&&\null+2(10^*,1)&&&&&&&&&&\\
SU(9)&Sp 4&19\{\bar{1}\}+3\{1^2\}+\{1^3\}+\{\bar{1^4}\}&4&(5^*,
4)+(10,4)+\null&5.0\times 10^{10}&1.93\times
10^{15}&0.233&0.4&-74/3&-41&3.1&12&4(1,1)+3(1,5)&1.1\times 10^{-3}\\
&&&&\null+17(5^*,1)+3(5,4)+\null&&&&&&&&&&\\
&&&&\null+(5,5)+(10,4)+\null&&&&&&&&&&\\
&&&&\null+(10,1)+(10^*,5)&&&&&&&&&&\\
SU(9)&SO(5)&13\{\bar{1}\}+2\{1^3\}+\{\bar{1^4}\}&4&(5^*,4)+(10,4)&
8.2\times 10^{10}&1.38\times
10^{15}&0.234&0.3&-28&-45&4.6&10&(1,1)+15(1,4)&1\times10^{-3}\\
&&&&\null+2(5,5)+10(5^*,1)
&&&&&&&&&&\\
&&&&+2(10,4)+(10^*,5)
&&&&&&&&&&\\
&&&&
+3(10^*,1)&&&&&&&&&&\\
SU(10)&SO(5)&10\{\bar{1}\}+3\{\bar{1^2}\}+2\{1^3\}&5&(5^*,5)+(10,5)+
\null
&6.6\times
10^{11}
&1.17\times 10^{15}&0.231&0.3&-194/3&-
48&3.5&5&10(1,5)+5(1,10)&6\times
10^{-4}\\
&&&&\null+2(5^*,5)+(10,5)+\null&&&&&&&&&&\\
&&&&\null+10(5^*,1)+2(5,10)+\null&&&&&&&&&&\\
&&&&\null+5(10^*,1)&&&&&&&&&&\\
SU(11)&SO(6)&12\{\bar{1}\}+4\{\bar{1^2}\}+2\{1^3\}&6&(5^*,6)+(10,6)+
\null
&3.3\times 10^{13}&1.37\times
10^{15}&0.234&0.1&-116/3&-33&13.6&2&14(1,6)+4(1,1)&2\times 10^{-4}\\
&&&&\null+3(5^*,6)+(10,6)+\null&&&&&&&&&&\\
&&&&\null+12(5^*,1)+2(5,15)+\null&&&&&&&&&&\\
&&&&\null+6(10^*,1)&&&&&&&&&&\\
\hline
\multicolumn{15}{c}{}\\
\multicolumn{15}{c}{
\mbox{\footnotesize
 Table 1: Examples of unified models with horizontal interactions, 
their parameters and second loop corrections from horizontal 
gauge boson coupling.
}} \\
\end{array}
$
}
}
\end{figure}
\def\thepage{}

\end{document}